\documentclass[conference]{IEEEtran}
\IEEEoverridecommandlockouts
\usepackage{cite}
\usepackage{amsmath,amssymb,amsfonts}
\usepackage{algorithmic}
\usepackage{graphicx}
\usepackage{textcomp}
\usepackage{svg}
\usepackage{xcolor}
\usepackage{url}
\usepackage{dblfloatfix}    

\def\BibTeX{{\rm B\kern-.05em{\sc i\kern-.025em b}\kern-.08em
    T\kern-.1667em\lower.7ex\hbox{E}\kern-.125emX}}
\begin{document}

\title{Deep Ultrasound Denoising Using Diffusion Probabilistic Models\\

}

\author{
    \IEEEauthorblockN{Hojat Asgariandehkordi\textsuperscript{1}, Sobhan Goudarzi\textsuperscript{2}, Adrian Basarab\textsuperscript{3}, Hassan Rivaz\textsuperscript{1}}
    \IEEEauthorblockA{\textsuperscript{1}Concordia University, Montreal, Canada}
    \IEEEauthorblockA{\textsuperscript{2}Sunnybrook Research Institute, Toronto, ON, Canada.}
    \IEEEauthorblockA{\textsuperscript{3}Univ Lyon, INSA-Lyon, Université Claude Bernard Lyon 1, CREATIS UMR 5220, U1294, F-69100, Villeurbanne, France.}
    \IEEEauthorblockA{hojat.asgariandehkordi@mail.concordia.ca, sobhan.goudarzi@sri.utoronto.ca}
        \IEEEauthorblockA{adrian.basarab@creatis.insa-lyon.fr, hrivaz@ece.concordia.ca}
}

\maketitle

\begin{abstract}
Ultrasound images are widespread in medical diagnosis for musculoskeletal, cardiac, and obstetrical imaging due to the efficiency and non-invasiveness of the acquisition methodology. However, the acquired images are degraded by acoustic (e.g. reverberation and clutter) and electronic sources of noise. To improve the Peak Signal to Noise Ratio (PSNR) of the images, previous denoising methods often remove the speckles, which could be informative for radiologists and also for quantitative ultrasound. Herein, a method based on the recent Denoising Diffusion Probabilistic Models (DDPM) is proposed. It iteratively enhances the image quality by eliminating the noise while preserving the speckle texture. It is worth noting that the proposed method is trained in a completely unsupervised manner, and no annotated data is required. The experimental blind test results show that our method outperforms the previous nonlocal means denoising methods in terms of PSNR and Generalized Contrast to Noise Ratio (GCNR) while preserving speckles.
\end{abstract}


\section{Introduction}

Using high frequency sound waves, ultrasound imaging visualizes body organs which assists medical practitioners in diagnosis applications\cite{r0}. As a low-cost, portable, and non-invasive modality, ultrasound is the second most popular medical imaging modality\cite{r1}. However, ultrasound images suffer from a low signal-to-noise ratio, which hinders their wider adoption. The construction of the ultrasound images is mainly divided into two steps. First, the excitation pulses stimulate a number of piezoelectric elements to transmit a particular beamshape into the medium. Then, the reflective waves are converted into channel radiofrequency (RF) data and converted into beamformed RF data. This RF data contains an additive Gaussian noise originating from various sources such as the sensors, acquisition card, etc. After envelope detection and log compression, the RF data is converted to the B-mode ultrasound images, which inherit the noise in the RF data (although the nonlinear operations change the noise statistics).

Many efforts have been dedicated previously to denoise ultrasound images, most notably nonlocal means (NLM)\cite{r2} and BM3D \cite{r3}. In NLM, the main idea is to denoise a pixel considering similar patches from a nonlocal neighborhood rather than just the neighboring pixels. In addition to direct denoising applications, the NLM has also been used recently in an inverse problem formulation for ultrasound beamforming\cite{r12}. In BM3D \cite{r3}, a collaborative filtering is used to effectively reduce noise in images. It exploits similarities between image patches to provide high-quality denoising results. While previous image-denoising methods have generally performed satisfactorily, they often exhibit a notable drawback. More specifically, these methods have shown a tendency to remove speckle noise, which can carry useful information for a more precise diagnosis.

With the remarkable progress of deep neural networks, innovative solutions have been proposed in medical image analysis, and ultrasound imaging is not an exception to this trend. For instance, deep learning has been used for reconstructing beamformed ultrasound data from raw radiofrequency channel data, utilizing both supervised and unsupervised approaches\cite{r8,r9,r10,r11}. Regarding that matter, our proposed method follows an unsupervised algorithm to learn how to eliminate different noise levels and retrieve the original image from its noisy counterpart. As a result, the proposed method is able to suppress the noise in such a way the speckle and the background texture are preserved.
\section{Background on DDPM}
The main idea behind DDPM \cite{r4} is learning how to remove the noise from an initial step through several short step sizes.
The vanilla diffusion model mainly includes two major steps: the forward process and the inverse process. 
\subsection{Forward process}
In the forward process, the objective is to add a weakened normal noise to the input image iteratively in such a way the image turns into pure noise during $T$ iterations.
More specifically, at each iteration $t$ of the forward process, a normal noise is multiplied by a coefficient that is less than one. This modified noise is then added to the image obtained from the previous step. This process can be conceptualized as a Markov chain, where each iteration depends on the outcome of the previous iteration. Mathematically, the forward process can be formulated as follows:
\begin{equation}
    q(x_t|x_{t-1}) = N_{x_t}(\sqrt{1-B_t} x_{t-1}, B_t I)
\end{equation}
where $B_t$ is a coefficient associated with the step $t$, and $I$ is the covariance matrix of the additive normal noise, which is assumed to be an identity matrix in the vanilla DDPM. Therefore, the output of each iteration (t) could be calculated as follows:
\begin{equation}
    x_t = \sqrt{1 - B_t} x_{t-1} + \sqrt{B_t} \varepsilon_{t-1},
\end{equation}
where $\varepsilon_{t-1}=N(0,I)$. If we assume $\alpha{_t} = 1 - B{_t}$, therefore:

\begin{align}
x_t &= \sqrt{\alpha_t} x_{t-1} + \sqrt{1-\alpha_t} \varepsilon_{t-1} \\&= \sqrt{\alpha_t \alpha_{t-1}} x_{t-2} + \sqrt{(1-\alpha_t) \alpha_{t-1}} \varepsilon_{t-2} \label{eq}
\end{align}
Following the above formula, we will have:
\begin{align}
x_t &= \sqrt{\overline{\alpha}_t} x_0 + \sqrt{1 -\overline{\alpha}_t} \varepsilon_0 \label{eq:example}
\end{align}
with
\begin{align}
\overline{\alpha} &= \alpha_t \cdot \alpha_{t-1} \cdot \alpha_{t-2} \cdot \ldots \cdot \alpha_0 
\end{align}

Hence, the noisy image in each arbitrary iteration $t$ can be reached directly from the original image ($x_0$).

\subsection{Inverse (backward) process}
Using the forward process, a number of noisy images are produced in each iteration whose noise distribution is determined for us. Consequently, an image-to-image deep neural network \cite{r11} (consisting of a time embedding module) can be trained to approximate the noise that was added to the image regarding the iteration $t$. After training the network to estimate the underlying noise in the images, an inverse process is established to iteratively retrieve the denoised image. The formulation of the inverse process can be described as follows:
\begin{align}
x_{t-1} &= \frac{1}{\sqrt{\alpha_t}} x_t + \frac{1 - \alpha_t}{\sqrt{1-\overline{\alpha}}} \varepsilon{_t} \label{eq:example}
\end{align}
where $\varepsilon{_t}$ is the approximated noise by the network giving the noisy image $x_t$ in step $t$. Also, $x_{t-1}$ is the denoised image that is fed to the network (as input) in the next iteration. It is worth noting that the inverse process is replicated in $T$ iterations to reach a denoised image.

Regarding the main idea that DDPM proposes, our contributions can be summarized as follows:

\begin{itemize}
    \item Constructing the forward and the inverse process with the associated parameters in the context of ultrasound denoising.
    \item Designing the image to image architecture consisting of a time-embedding module to impose the time steps to the model.

\end{itemize}

\section{Methods}
To use the DDPM for the denoising application in ultrasound images, we utilized the same forward and inverse process explained in the previous section, and the parameters $T$, $B_t$, and $I$ were considered as $300$, $1/300$, and $1$, respectively. For our image-to-image network, we designed a U-Net-based network \cite{r5} alongside a time embedding module to inform the network about the time step $t$. More information about the designed architecture is shown in Fig. 1.
\begin{figure*}
  \centering
  \includegraphics[width=0.8\textwidth]{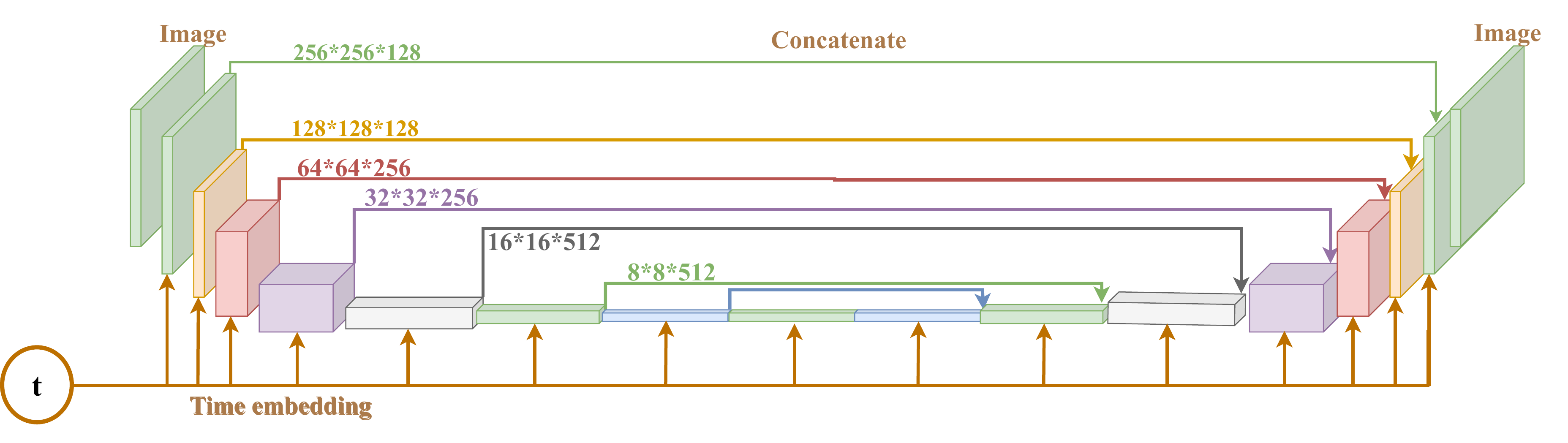}
  \caption{The proposed architecture.}
  \label{fig:example}
\end{figure*}

\section{Results and Discussions}
In this section, we will first explain the datasets were used in this study. Subsequently, we will delve into the training strategy employed, and finally, the experimental results are going to be reported in detail.
\subsection{CIFAR10 dataset \cite{r6}}
The CIFAR-10 dataset contains 60000 images in 10 classes, each having 6000 color images of size 32x32. The dataset has been separated into 50000 training and 10000 test images. The dataset is divided into six batches, with five training batches that contain 10000 randomly ordered images (for each batch) and one test batch that consists of 10000 images randomly selected from each category. While the training batches include images from all ten classes, each batch might contain a different number of images from each group. Nonetheless, collectively, the training batches contain precisely 5000 images from each category.
\subsection{CUBDL dataset \cite{r7}}
The CUBDL dataset includes a set of raw Radio Frequency (RF) data from different transmission angles that were collected for three different tasks. In this study, 1520 images (corresponding to the task1 of the CUBDL dataset) are reconstructed (a single image was reconstructed for each transmission angle separately). The beamformed images are split into train and test sets with a ratio of 15 percent.
\subsection{Experimental settings}
The proposed network is pre-trained on CIFAR10 dataset \cite{r6} and subsequently, fine tuned on the CUBDL dataset \cite{r7}. For all of the steps, the hyperparameters are determined as explained in Table I.

\begin{table}[h]
\caption{Hyperparameters for the model.}
\centering
\begin{tabular}{lcc}
  \hline
  Hyperparameter & Amount \\
  \hline
  Image Size & 256*256 \\
  Batch Size & 32 \\
  Optimizer & Adam \\
  Learning rate for Stanford Cars dataset & 0.001 \\
  Learning rate for CUBDL dataset & 0.0004 \\
  LRscheduler & StepLR(gamma=0.3) \\
  Epochs & 200 \\
  Loss Function & Mean Squared Error \\
  \hline
\end{tabular}
\end{table}

Pre-training the network on CIFAR10\cite{r6} dataset was done in two steps; first, we trained the network from scratch on images with the size 64*64. Then, using the initial weights of the first step, the network was fine tuned on the images of size 256*256. When it comes to fine-tuning on the CUBDL dataset, the network was fine tuned on images of size 256*256 using initial weights of the pre-training step. All of the steps were implemented on an NVIDIA Tesla A 100 (40 GB RAM). The training time on CIFAR 10 dataset was around 8  hours and 38 minutes for image size 256*256. As for the CUBDL dataset, the training time decreased to around three hours. The test errors (using L.1 norm loss function) of the network on both datasets are reported in Table II.
\begin{table}
    \centering
    \caption{Comparison of error rates in the first and last epochs}
    \label{tab:error_rates}
    \begin{tabular}{lcc}
        \hline
        Dataset & Error in the first epoch & Error in the last epoch \\
        \hline
        Cifar 10 & 2.75 & 0.05 \\
        CUBDL & 2.64 & 0.039 \\
        \hline
    \end{tabular}
\end{table}

\subsection{Peak Signal to Noise Ratio (PSNR) metric}
Peak signal-to-noise ratio (PSNR) is the ratio between the maximum possible power of a signal and the power of corrupting noise that affects the fidelity of its representation. Because many signals have a very wide dynamic range, PSNR is usually expressed in a logarithmic quantity using the decibel scale. PSNR is commonly used to quantify the reconstruction quality of images and video subject to lossy compression. Given a noise-free $m\times n$ monochrome image I and its noisy approximation K, PSNR is defined as:

\begin{equation}
MSE = \frac{1}{{MN}} \sum_{i=0}^{M-1} \sum_{j=0}^{N-1} [I(i,j) - K(i,j)]^2
\end{equation}
\begin{equation}
PSNR = 10 \log\left(\frac{{\text{{MAX}}(I)}}{{\text{{MSE}}}}\right)
\end{equation}
where $I$ is the original image, and $K$ is the denoised version.
\subsection{Results}
To show the superiority of the proposed method as compared to the most prevalent denoising algorithms, the results of BM3D \cite{r7} and NLM \cite{r6} methods in different noise levels are presented. To this end, we added a normal noise through the forward process on test images for $T= 10, 20$ iterations. Then, the noisy images were denoised by each individual denoising method. The results of PSNR and GCNR are reported in Tables III and IV, respectively. As can be deducted in Table III, our method surpasses the other two methods with a sensible margin for both $T=10$ and $20$, which means the proposed method is more capable of vanishing the existing noise in the image while keeping it close to the original image. Regarding the GCNR results in Table IV, the results for $T=10$ were very similar among the methods. However, for $T=20$, the proposed method achieved a better GCNR value compared to the other two methods.

\begin{table}[h]
\caption{PSNR values for different techniques at different $T$ values}
\centering
\begin{tabular}{lcc}
  \hline
  Technique & $T=10$ & $T=20$ \\
  \hline
  Noisy Image & 22.5 & 17.8 \\
  NLM & 22.3 & 22.03 \\
  BM3D & 22.3 & 22.4 \\
  OURS & 24.73 & 23.6 \\
  \hline
\end{tabular}
\end{table}

\begin{table}
    \centering
    \caption{Comparison of GCNR(\%) for different methods for $T=20$}
    \label{tab:gcnr}
    \begin{tabular}{lcc}
        \hline
        Method & GCNR(\%) \\
        \hline
        NLM & 91.6 \\
        BM3D & 90.4 \\
        OURS & 92.6 \\
        \hline
    \end{tabular}
\end{table}

Apart from the numerical results, we have conducted some visualization experiments on the test images. In Fig. 2, the noisy images and the output of each method are illustrated. It is worth noting the that original image is shown in Fig. 2. has been reconstructed using the (Delay And Sum) DAS method and on the channel data of 75 different beam angles.

\begin{figure*}
  \centering
  \includegraphics[width=1\textwidth,height=0.4\textwidth]{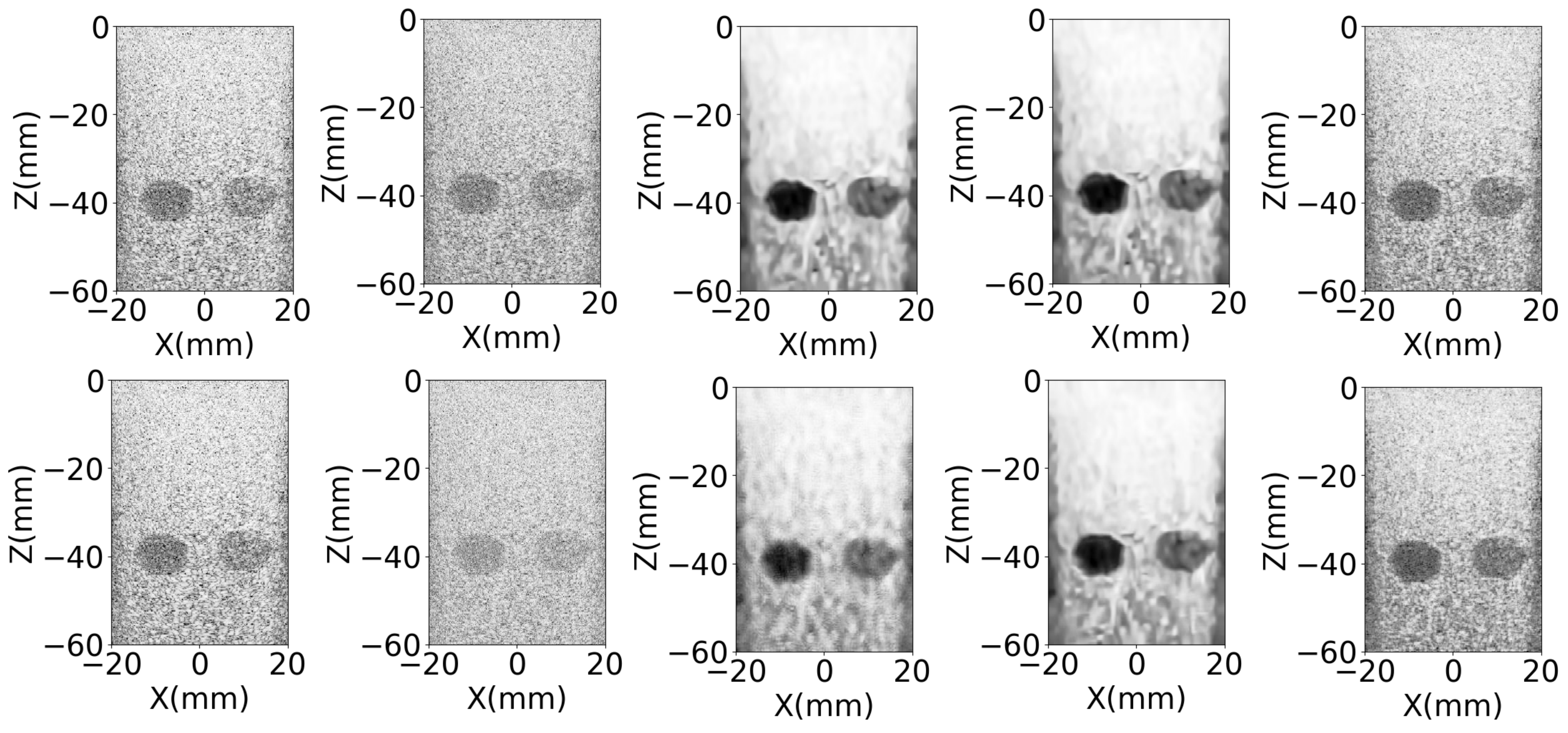}
  \caption{Experimental phantom results. From left to right: Original image, Noisy image, NLM, BM3D, and Our method. The first row refers to $T=10$ and the second one shows the results of $T=20$. The original image is from the test set of the CUBDL dataset, and was collected by the  LA-533 transducer at a center frequency of $8$ MHz.}
  \label{fig:example}
\end{figure*}

Based on the illustrated images in Fig. 2, although NLM and MB3D can provide denoised images with distinguishable cysts, the resulting outputs are getting blurred with respect to the number of iterations $T$. On the other hand, our proposed method aims to keep the background texture while denoising the image. This advantage could be helpful in diagnosis applications in which background plays an essential role in detecting a particular disease.

\section{Conclusion}
In this paper, noisy ultrasound images were represented as one of the major challenges in medical diagnosis. Regarding that, stemming from denoising diffusion probabilistic models, a denoising method is proposed that retrieves the denoised image by passing the noisy image through an iterative inverse process. The comparison results show that our method outperforms other denoising methods in both quantitative and qualitative experiments.

\section*{Acknowledgement}
We thank Natural Sciences and Engineering Research Council of Canada (NSERC) for funding.
\bibliographystyle{unsrt}
\bibliography{name}
\end{document}